\documentclass[twocolumn,english,aps,prb,showpacs,superscriptaddress,amssymb,amsfonts]{revtex4}
\usepackage[T1]{fontenc}
\usepackage[latin9]{inputenc}
\usepackage{amsmath}
\usepackage{graphicx}
\usepackage{amssymb}
\usepackage{color}
\makeatletter
\newcommand{\thickhline}{\noalign{\hrule height 0.8pt}}
\newcommand{\be}{\begin{equation}}
\newcommand{\ee}{\end{equation}}                  
\newcommand{\bea}{\begin{eqnarray}}
\newcommand{\eea}{\end{eqnarray}}
\newcommand{\beas}{\begin{eqnarray*}}
\newcommand{\eeas}{\end{eqnarray*}}

\newcommand{\upa}{\uparrow}
\newcommand{\dna}{\downarrow}

\makeatother

\usepackage{babel}

\makeatother

\usepackage{babel}

\begin{document}
 
 \title{Quantum Phase Transition between Integer Quantum Hall States of Bosons}

\author{Tarun Grover}

\affiliation{Kavli Institute for Theoretical Physics, University of California, Santa Barbara, CA 93106, USA}

\author{Ashvin Vishwanath}
\affiliation{Department of Physics, University of California,
Berkeley, CA 94720,
USA}

\begin{abstract}
Bosonic Integer Quantum Hall (IQH) phases are a class of symmetry protected topological (SPT)  phases that, similar to the fermionic IQH states,  support  a quantized Hall conductance. They, however, require interactions for their realization. Here we study quantum Hall plateau transitions between  a trivial insulator and a bosonic IQH phase,  in a  clean system.  Generically, we find an intervening superfluid phase. The presence of additional symmetries, however, can potentially lead to a direct transition between these phases. We employ a fermionic parton description that captures both the insulating phases as well as the transition between them. 
The critical theory is massless QED-3 with $N_f=2$ fermion flavors. The fermions have a surprisingly simple interpretation - they are vortices of the superfluid. Therefore the universal conductivity at the transition, assuming it is continuous, equals the universal {\em resistivity} of Dirac fermions in QED-3. We also briefly comment on sigma model descriptions of the transition and the surface states of related 3D SPT phases.

\end{abstract}

\maketitle
\tableofcontents
 \section{Introduction} \label{sec:intro}
Quantum phase transitions \cite{SachdevBook} beyond the standard Landau-Ginzburg-Wilson paradigm provide an exciting opportunity to study universal phenomena in  strongly correlated systems. When a transition occurs between phases that differ at the level of topology, rather than symmetry, the critical properties  naturally lie beyond the standard paradigm. The best studied example involves transitions between quantum Hall states  \cite{sondhireview, wen1993, chen1993, ludwig1994, sachdev1998, senthil1999, read2000, wen2000, maissam2012}, or quantum spin-liquids\cite{chubukov1994}. In principle, non-Landau transitions are also possible even when the phases on the either side of the transition are conventional symmetry broken states, examples of which were considered in frustrated magnets and lattice boson systems \cite{senthil2004, balents2005,sandvik2007, melko2008}, and were dubbed `deconfined quantum critical points'. Here, the unusual physics arises since defects in the order parameter field on either side of the phase transition carry non-trivial quantum numbers. In this paper, we provide a new example of deconfined quantum criticality relevant for  strongly interacting bosonic systems.

Motivated by the recent discovery of (fermionic) $\mathbb{Z}_2$ topological insulators \cite{hasanrev, qirev, moorerev}, it was recently realized that bosons or spins can realize topological phases in various dimensions, that however do not have topological order and hence no exotic excitations in the bulk. An early example is the Haldane chain \cite{haldane1983}, and generalizations thereof in one dimension \cite{gu2009, fidkowski2011,chen2011_1d,schuch2011,turner} where symmetry protects special edge modes that define the phase. Generalizations of such phases in various dimensions have been an active area of theoretical research \cite{chen2011, chen2011_edge, kitaevtalk, levin2012, lu2012, levinsenthil2012, senthil2012}. Akin to the fermionic topological insulators, these new phases have the interesting feature that in the presence of a certain symmetry (for example, particle number conservation), they support protected edge states. However unlike fractional quantum Hall states, there are no exotic anyon excitations \cite{footnoteE8}. If the symmetry is broken, either explicitly or spontaneously, the system is indistinguishable from a featureless Mott insulator. A crucial difference between the bosonic SPT phases and the fermionic topological insulators is that the latter can be realized in a non-interacting system, while the former necessarily requires interactions, since a system of non-interacting bosons will just bose-condense. 

One of the simplest example of a bosonic SPT phase in 2D is the bosonic IQH state \cite{chen2011,lu2012,levinsenthil2012}, which supports gapless edge states and a Hall effect quantized to {\em even} integer multiples of the quantum of conductance\cite{lu2012,levinsenthil2012}, i.e. $\tilde{\sigma}_{xy}=2n$ in dimensionless units. In Ref. \onlinecite{levinsenthil2012} a particularly simple model Hamiltonian was proposed to realize this phase - comprising of a two component fluid of bosons in the lowest Landau level at $\nu_{tot}=2$, with contact interactions. It was also proposed that  cold atoms in artificial gauge fields could eventually realize such Hamiltonians\cite{levinsenthil2012}. 

In this paper we study quantum phase transitions between two different bosonic IQH states, which furnishes an example of a quantum critical point separating ``symmetry protected topological'' (SPT) phases \cite{ chen2011_edge, chen2011, kitaevtalk, levin2012, lu2012, levinsenthil2012, senthil2012}. We find that these transitions furnish a new class of non-Landau phase transitions. To achieve this goal, we devise a fermionic parton construction for the bosonic SPT phases that is ideally suited to describe the phases as well as the transitions between them. 

We find that generically, the two SPT phases are separated by an intermediate superfluid phase, and the insulator-superfluid transition is the conventional one based on the Landau order parameter distinction between these phases. 

However, in the presence of lattice inversion and a discrete internal symmetry (analogous to charge conjugation symmetry), the two SPT phases can be potentially connected by a second order transition. The critical theory corresponds to two flavors of Dirac fermions coupled to a non-compact $U(1)$ gauge field.  
\bea
\mathcal{L}_{critical}& = &  \sum_{\alpha = +/-} \overline{f}_{\alpha} [\gamma_\mu(-i \partial_\mu  - a_{\mu} ) ] f_{\alpha} + \frac{(\partial_\mu a_{\nu}-\partial_{\nu}a_{\mu})^2 }{2g^2} \nonumber \\
& &  - \frac{1}{2\pi}\epsilon^{\mu\nu\lambda} A_\mu \partial_\nu a_\lambda  -  \frac{1}{4\pi} \epsilon^{\mu\nu\lambda} A_\mu{\partial_\nu A_\lambda}  
\label{qedintro}
\eea
where $A$ represents the external electromagnetic field and $\gamma_\mu$ are a set of three $2\times 2$ Dirac matrices. The first two terms are referred to as  the QED-3 action, with $N_f=2$ flavors of fermions. Satisfyingly, here the fermions and the gauge field have a simple physical interpretation: the third term implies that the flux of the gauge field $a$ is just the boson density.  And the minimal coupling of fermions to $a$ implies that they are vortices.

The question of stability of QED-3  has been an active topic of research \cite{appelquist}.  Fluctuations could lead to spontaneous symmetry breaking and an energy gap to fermions, which is termed chiral symmetry breaking. While large $N_f$ is expected to be stable, spontaneous symmetry breaking occurs below a certain critical $N_f$, whose precise value is debated. For $N_f=2$ (two component Dirac fermions) of interest here, the model of Ref.\onlinecite{Hands} finds chiral symmetry breaking but with a small condensate, while in other approaches the possibility of a chiral symmetric phase has been proposed \cite{anisotropicQED}. Here, in the interest of simplicity, we assume fluctuations do not induce chiral symmetry breaking. If they indeed do - the continuous transition we discuss will presumably be replaced with an intervening superfluid that breaks lattice symmetries. Nevertheless, the QED-3 description provides the right variables to discuss all relevant phases within a unified formulation. Moreover,  if the correlation length is sufficiently long, the QED-3 description will be quantitatively accurate at intermediate length scales.

It is interesting to contrast this with the quantum Hall transition of free fermions. In the clean limit, the critical theory is a free Dirac fermion. Here, the critical theory features a pair of Dirac fermions, which, being vortices, are coupled to a gauge field. Via vortex-particle duality, the universal conductivity at the transition is shown to be simply the {\em resistivity} of the fermions, i.e. $\tilde{\sigma}_{xx}=1/\tilde{\sigma}_f$. The last term in Eqn. \ref{qedintro} implies $\tilde{\sigma}_{xy}=1$ at the transition. Disorder is expected to change the nature of these transitions, as with free fermions, and is a challenging open problem.

A different approach to this transition can be constructed based on a Chalker-Coddington network model picture \cite{ChalkerCoddington,senthil2012} or on the sigma model description of SPT phases\cite{chen2011}. These lead to variants of an O(4) model with a topological $\theta=\pi$ term, which is also related to theories of deconfined criticality, discussed in the context of frustrated magnets\cite{senthil2004, tanaka2005, senthil2006}.  Similar theories arise in describing surface states of 3D SPT phases, where time reversal symmetry is present in addition to charge conservation \cite{senthil2012}. We briefly discuss these connections and possible caveats.

\section{Fermionic parton description of Bosonic IQH phases} \label{sec:parton}

Our goal is to describe quantum phase transitions in bosonic symmetry protected topological (SPT) phases. We focus on SPT phases  with two species of bosons, both at the filling of one-particle per site. The underlying symmetry that protects the edge modes corresponds to total particle number conservation  of bosons \cite{chen2011, lu2012, levinsenthil2012} and thus, we call these systems bosonic integer quantum Hall (IQH) phases. We will ultimately allow the bosons of different species to mix, so that the only remaining symmetry is the total boson number. The classification for this class of SPT phases is $\mathbb{Z}$ \cite{chen2011, lu2012}, that is, there are an infinite number of SPT phases. Physically, the different phases in this class are labeled by their Hall conductance, which is an {\em even} integer \cite{lu2012,levinsenthil2012}. 
\begin{equation}
\sigma_{xy}=2n\frac{Q^2}{h}
\end{equation}
where $Q$ is the fundamental boson charge and $h$ is Planck's constant. We  describe conductivities in term of the dimensionless constants $\tilde{\sigma}$ where  $\sigma=\tilde{\sigma} \frac{Q^2}{h}$. Henceforth set $Q=\hbar=1$ and in these units $\sigma=\tilde{\sigma} /2\pi$.

One possible approach to describe such a phase is to use the non-linear sigma model approach introduced in Ref. \cite{chen2011}. Within this approach, the trivial SPT phase is described by an $O(4)$ model without a theta term while the non-trivial SPT phase that has a Hall conductance of $\sigma_{xy}=2$ is described by an $O(4)$ model with a theta term at $\theta = 2\pi$.  It is therefore not unreasonable to conjecture that the transition between these phases is controlled by a critical point with $\theta=\pi$, which is consistent with the observation in Ref.\cite{cenke2011} that this would lead to a non-degenerate ground state. Furthermore, in Sect IV B of  Ref.\onlinecite{senthil2012}, a network model construction of the transition between these two phases was analyzed. An $O(4)$ model at $\theta = \pi$ was obtained, which was then reduced to a model with just the physical $U(1)$ symmetry of charge conservation. Here we will offer a complementary approach, utilizing fermionic patrons, which is also of interest in its own right. Eventually we will try to build a correspondence to these sigma models with topological terms, as well as theories of deconfined criticality in frustrated magnets.

In Ref.\cite{lu2012}, an alternative description of SPT phases in terms of multi-component Chern-Simons theory was developed. Here, one considers multiple components of bosons indexed by $I=1,2,\dots N$. The currents of these different species are represented by a gauge field: $j^\mu_I=\epsilon^{\mu\nu\lambda}\partial_\nu a^I_\lambda/2\pi$. Then, SPT phases with a conserved charge are described by the action:
\begin{equation}
{\mathcal L}=\frac{K_{IJ}}{4\pi}\epsilon^{\mu\nu\lambda}a^I_\mu\partial_\nu a^J_\lambda - \frac{q_I}{2\pi} \epsilon^{\mu\nu\lambda}a^I_\mu\partial_\nu A_\lambda
\end{equation}
where the $K$ matrix is a symmetric, unimodular (i.e. det${\bf K}=1$) matrix with integer entries and even integer diagonal entries. The coupling to an external electromagnetic field is accomplished via the charge vector $q_I$. The quantized Hall conductance is given by:

$$
\tilde{\sigma}_{xy}={\bf q}^T {\bf K}^{-1} {\bf q}
$$

The integer quantum Hall states of bosons, labeled by $n$,  are described by the simple forms:
\begin{equation}
K_n= \begin{bmatrix} 0 & 1 \\ 1 & 2(1-n) \end{bmatrix};\,q= \begin{bmatrix} 1 \\1\end{bmatrix}
\end{equation}
and have $\tilde{\sigma}_{xy}=2n$.  The simplest nontrivial phase with $n=1$ has:

\begin{equation}
K_1= \begin{bmatrix} 0 & 1 \\ 1 & 0 \end{bmatrix};\,q= \begin{bmatrix} 1 \\1\end{bmatrix}
\end{equation}
described by the topological theory:

\begin{equation}
2\pi {\mathcal L}=\epsilon^{\mu\nu\lambda}a_{1\mu}\partial_\nu a_{2\lambda} -  \epsilon^{\mu\nu\lambda}(a_{1\mu}+a_{2\mu})\partial_\nu A_\lambda
\end{equation}

In the rest of the paper, we will denote $\epsilon^{\mu\nu\lambda}a_{I\mu}\partial_\nu a_{J\lambda}$ by 
$\epsilon a_{I}\partial a_{J}$. Note that  $\epsilon a_{J}\partial a_{I} = \epsilon a_{I}\partial a_{J}$.

\subsubsection{Wavefunction and Parton Construction} \label{sec:wfnparton}
The multi-component Chern-Simons description of the SPT phases motivates a parton construction for these phases, similar to that for conventional fractional quantum Hall phases. Let us first describe this approach using lowest Landau level wavefunctions. It is well known that the wave function corresponding to a two component quantum Hall state $\psi(z^\uparrow_i,z^\downarrow_i)$, where the coordinates of the the first (second) species is $z^{\uparrow(\downarrow)}_i$, and $i-1\dots M$ label the number of particles in each species, is:  
\begin{eqnarray}
\psi(z^\uparrow_i,z^\downarrow_i)&=& \prod_{i<j} (z^\uparrow_i-z^\uparrow_j)^{K_{11}} (z^\downarrow_i-z^\downarrow_j)^{K_{22}}\nonumber\\
&& \prod_{i,j} (z^\uparrow_i-z^\downarrow_j)^{K_{12}}
\end{eqnarray}
where we have ignored the exponential factors. Note, the simplest nontrivial IQH phase of bosons is then given by the wave function: 
\begin{equation}
\psi_1(z^\uparrow_i,z^\downarrow_i)={\mathcal J(z^\uparrow_i,z^\downarrow_i)}(\prod_{i,j} (z^\uparrow_i-z^\downarrow_j)
\label{SPTwavefn}
\end{equation}

where ${\mathcal J(z^\uparrow_i,z^\downarrow_i)}$ is a real number that describes a Jastrow factor which ensures a homogeneous phase. Now we describe a parton construction that yields the same state. 

So let us consider two species of bosons $b_\upa$ and $b_\dna$, each at filling $\nu =1$ in the lowest Landau level. One way to describe the nontrivial IQH phase is  to use the following parton construction:

\beas
b_\upa & = & \psi_\upa \,\psi_0 \\
b_\dna & = & \psi_\dna \,\psi_0 \\
\eeas

where the partons $\psi_\upa \,\psi_\dna, \,\psi_0$ are fermionic, and are placed in filled Landau levels. We assign a charge of $+2$ to $\psi_0$ and $-1$ to $\psi_1,\, \psi_2$.  The partons $\psi_0$ sees a magnetic field of the same sign but twice the strength, the other two patrons $\psi_\upa,\,\psi_\dna$ see a reversed field.  There are twice as many $\psi_0$ particles, as those of $\psi_1$ (or $\psi_2$). Thus, they individually fill Landau levels. Then, the projection implied by the construction above, requires one to identify the coordinates of the $\psi_0$ particles with those of both $\psi_\upa$ and $\psi_\dna$, which in turn are identical to the coordinates of the two species of bosons. Putting this together we get:
\begin{equation}
\psi_{parton} =  \prod_{i<j} |z^\uparrow_i-z^\uparrow_j|^2|z^\downarrow_i-z^\downarrow_j|^2 \prod_{i,j} (z^\uparrow_i-z^\downarrow_j)
\end{equation}
which, is of the same form as the Eqn. \ref{SPTwavefn}. Thus we are able to describe the topological IQH phase\cite{levinsenthil2012,jain93}. In order to describe the trivial phase within the same formalism, and hence a transition, we make two extensions. First, we introduce a slightly more complicated parton construction and second we assume the presence of a lattice. The second assumption allows us to consider bands with Chern number which can change as we tune parameters. This will allow us to discuss transitions between SPT phases. We note that the above parton construction also arises in the context of certain non-abelian quantum Hall states \cite{maissamwen2012}.

\subsubsection{Parton Description and Field Theory} \label{sec:partonfield}
So let us consider two species of bosons $b_\upa$ and $b_\dna$, each at filling $\nu =1$ on a lattice. One way to describe the trivial as well as the IQH phase within the same formalism is 
to use the following parton construction:

\beas
b_\upa & = & \psi_\upa \,\psi_0 \\
b_\dna & = & \psi_\dna \,\psi_0 \, f_1 \,f_2\\
\eeas

where the partons $\psi_\upa \,\psi_\dna, \,\psi_0, f_1, f_2$ are all fermions. As with any parton construction,  the above redefinition of the microscopic fields $b_\upa, b_\dna$ is redundant which leads to internal gauge field degrees of freedom that couple to the partons. In particular, the above construction leads to three abelian gauge fields $a_1, a_2, a_3$ being coupled to the partons and we assign the gauge charges shown in the first three columns of the Table \ref{table1} to the partons. The bosons $b_\upa, b_\dna$ are both charged under the (non-dynamical) electromagnetic field $A_c$ with unit charge, and we account this by assigning a gauge charge of  unity each to the partons $\psi_\upa$ and $\psi_\dna$. If one prefers, one may alternatively assign the partons the same electromagnetic charge as in the case of lowest Landau construction mentioned above. All the gauge-invariant quantities of course remain unchanged under any such re-assignment.

We now demonstrate that the assignment of gauge charges and Chern numbers in Table \ref{table1} yields the desired phases. We also include superfluid phase in the table since it will be needed in the discussion that follows.

\begin{table}
\begin{tabular}{ |c|c | c | c| c|c|c| c|}
   \hline                      
  & $a_1$ & $a_2$ & $a_3$ & $A_c$ & Chern $\# $ in & Chern $\#$ in &  Chern $\#$ in \\ 
& & & & & Trivial Phase & IQH Phase & Superfluid \\ \hline
$\psi_\upa$ & 1 & 0 & 0 & 1 & 1 & 1 & 1 \\ \hline
$\psi_\dna$ & 1 & 1 & 0 & 1 & 1 & 1 & 1 \\  \hline  
$\psi_0$ & -1 & 0 & 0 & 0 & -1 & -1 & -1 \\  \hline  
$f_1$ & 0 & -1 & 1 & 0 & -1 & -1  & -1 \\  \hline  
$f_2$ & 0 & 0 & -1 &0 & -1 & 1 & 0\\  \hline  
\end{tabular}
\caption{The gauge charges and the Chern number assignments for various partons in the three distinct phases studied in this paper.}
\label{table1}
\end{table}

\subsection{Integer Quantum Hall Phase of Bosons} \label{sec:topospt}

Let us first consider a phase where $\psi_\upa, \psi_\dna, f_2$ are in a band with Chern number 1 while $\psi_0, f_1$ are in a Chern number -1 band (see Table I). 
Defining the parton currents $j^I = \frac{1}{2\pi} \nabla \times \alpha^I$ where $I$ labels the parton (i.e. $I = \upa, \dna$, etc.), the low energy theory is 
given by 

\bea 
\mathcal{L} &  = &  -\frac{\epsilon}{4\pi}(\alpha^{\upa} \, \partial \alpha^{\upa} + \alpha^{\dna} \, \partial \alpha^{\dna} - \alpha^0 \, \partial \alpha^0 \nonumber \\ 
& & - \alpha^1 \, \partial\alpha^1 + \alpha^2 \, \partial\alpha^2) \nonumber + \mathcal{L}_{constraint} \label{nontrivL1}
\eea

where 

\bea
\mathcal{L}_{constraint} & = & \frac{\epsilon}{2\pi} \left[ a_1 \, \partial(\alpha^\upa + \alpha^\dna - \alpha^0) +  a_2 \, \partial(\alpha^\dna - \alpha^1) \right. \nonumber \\
& &  + \left. a_3 \, \partial(\alpha^1 - \alpha^2) +  A_c \, \partial(\alpha^\upa + \alpha^\dna) \right]  \label{eq:constr}
\eea

Solving the constraint equations by integrating out the internal gauge fields $a_i$, one obtains

\be 
\mathcal{L} = \frac{\epsilon}{4\pi}(\alpha^{\upa} \, \partial\alpha^{\dna} + \alpha^{\dna} \, \partial\alpha^{\upa}) + \frac{\epsilon}{2\pi} A_c \, \partial(\alpha^\upa + \alpha^\dna) \label{nontrivL2}
\ee

In the $K$-matrix formulation of SPT phases, this phase corresponds to $K = \begin{bmatrix} 0 & 1 \\ 1 & 0 \end{bmatrix}$ with a charge vector $q^T = [1\,\,\,1]$ and hence $\tilde{\sigma}_{xy}=2$. 

In passing, we note that putting both $f_1$ and $f_2$ in Chern number 1 band, while keeping the Chern numbers of other partons same as here, one obtains a IQH state with  $\tilde{\sigma}_{xy}=4$.
%

\subsection{Trivial Insulator} \label{sec:trivspt}

Next, consider a different phase where $f_2$ is in a Chern number -1 band while all the other partons have the same Chern number as before (Table \ref{table1}). The Lagrangian is given by  

\bea 
\mathcal{L} &  = &  -\frac{\epsilon}{4\pi}(\alpha^{\upa} \, \partial\alpha^{\upa} + \alpha^{\dna} \, \partial\alpha^{\dna} - \alpha^0 \, \partial\alpha^0 \nonumber \\ 
& & - \alpha^1 \, \partial\alpha^1 - \alpha^2 \, \partial\alpha^2) \nonumber + \mathcal{L}_{constraint} \label{trivL1}
\eea

where $\mathcal{L}_{constraint}$ is same as before (since the gauge charge assignment for the partons remains unchanged).

Proceeding as before, we integrate out the internal gauge fields $a_i$ and obtain the following Lagrangian:

\be 
\mathcal{L} = \frac{\epsilon}{4\pi}(2 \alpha^{\dna} \, \partial\alpha^{\dna} + 2\alpha^{\upa} \, \partial\alpha^{\dna} + \frac{1}{2\pi} A_c \, \partial(\alpha^\upa + \alpha^\dna)) \label{trivL2}
\ee
 
which is the $K = \begin{bmatrix} 0 & 1 \\ 1 & 2 \end{bmatrix}$ theory, with the same charge vector as before, $q^T = [1\,\,\,1]$ . The charge Hall conductance is given by  $\sigma^c_{xy} = [1\,\, 1] K^{-1} [1\,\, 1]^T = 0$, as expected. 
We also note that there is no Meissner term generated for $A_c$, verifying that this phase is indeed an insulator (and not a superfluid).

\subsection{Superfluid Phase} \label{sec:sf}

The superfluid phase is obtained when $f_2$ is in Chern number zero band while everything else remains unchanged. In this case, the low energy theory after integrating out partons is given by:

\bea
\mathcal{L}_{SF} & = & \frac{\epsilon}{4\pi} \left[ (a_1 + A_c) \, \partial(a_1 + A_c) \right. \nonumber \\
 & &  + (a_1 + a_2 + A_c) \, \partial(a_1 + a_2 + A_c) \nonumber \\
& & - \left. a_1 \, \partial a_1 - (a_2 - a_3) \, \partial(a_2 - a_3) \nonumber \right]
\eea

Integrating out $a_3$ and $a_2$ successively, leads to 

\be
\mathcal{L}_{SF} = \frac{\epsilon}{4\pi} (-2a_1 \, \partial A_c + A_c \, \partial A_c)
\ee

which is clearly a superfluid phase since the integration of $a_1$ generates a Meissner term for $A_c$. An alternative way to reach the same conclusion is to note that the   Chern number zero for the parton $f_2$ enforces the  
constraint $\alpha^2 = 0$  (recall the definition $j^I = \frac{1}{2\pi} \nabla \times \alpha^I$ for the parton current). This is because parton $f_2$ is fully localized in the real space. Following the other constraint equations (see Eqn.\ref{eq:constr}), this implies that $\alpha^1 = \alpha^\dna = 0$ as well. This leaves us with

\be
\mathcal{L}_{SF} =  \frac{\epsilon}{2\pi} A_c \, \partial \alpha^\upa
\ee

which again leads to a Meissner term for $A_c$.

\section{Phase Transitions and Critical Theories} \label{sec:spt_crit}
\begin{figure}[tb]
\centerline{
\includegraphics[width=240pt, height=180pt]{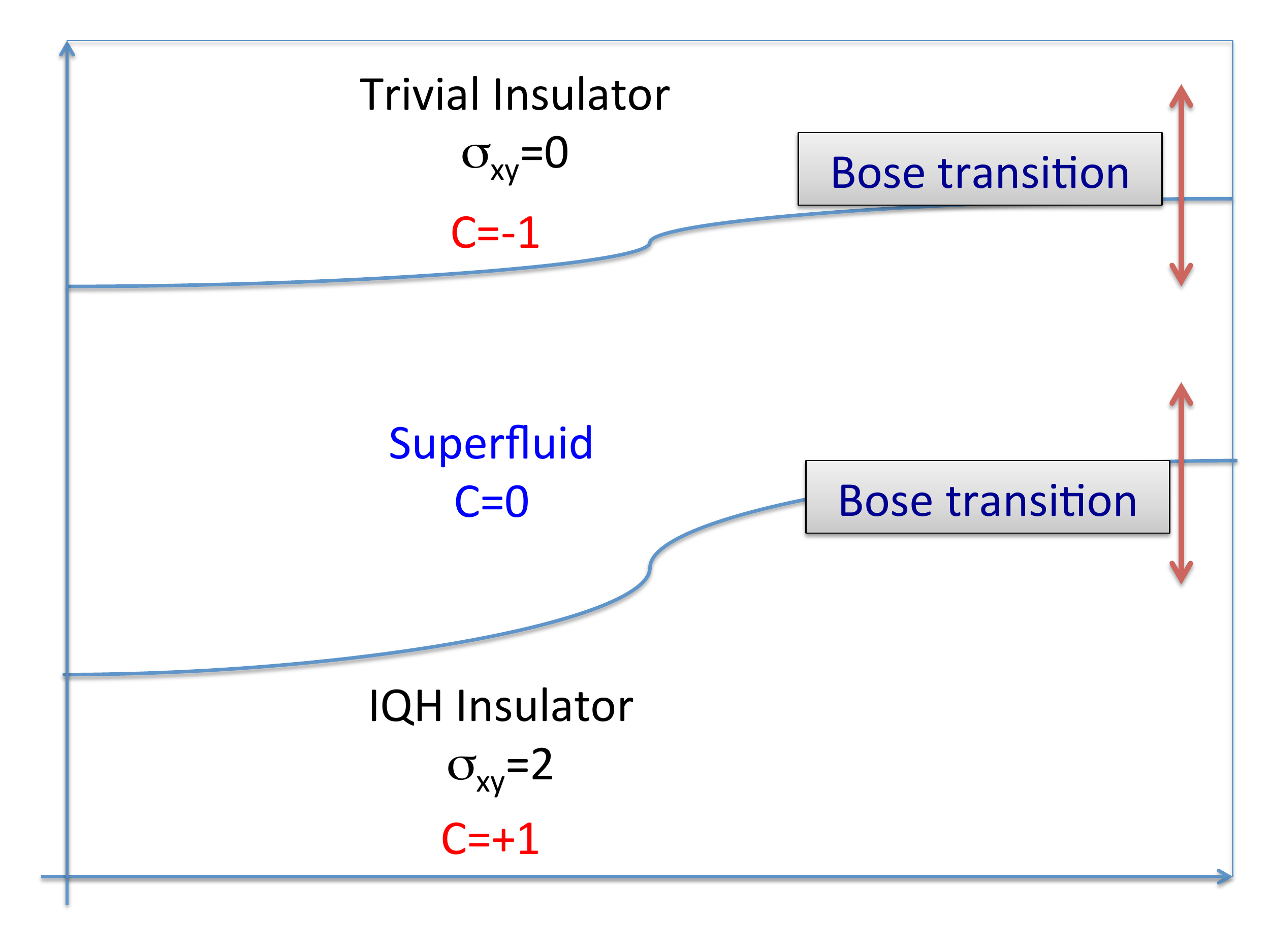}}
\caption{The generic phase diagram for the problem with no additional symmetry beyond charge conservation. The two superfluid-insulator phase transitions involve change of Chern number 1 for the parton $f_2$ (see the main text). These correspond to the {2+1D Bose-Einstein-Condensation transitions.}}
\label{fig:phasedianosym}
\end{figure}

\begin{figure}[tb]
\centerline{
\includegraphics[scale=0.35]{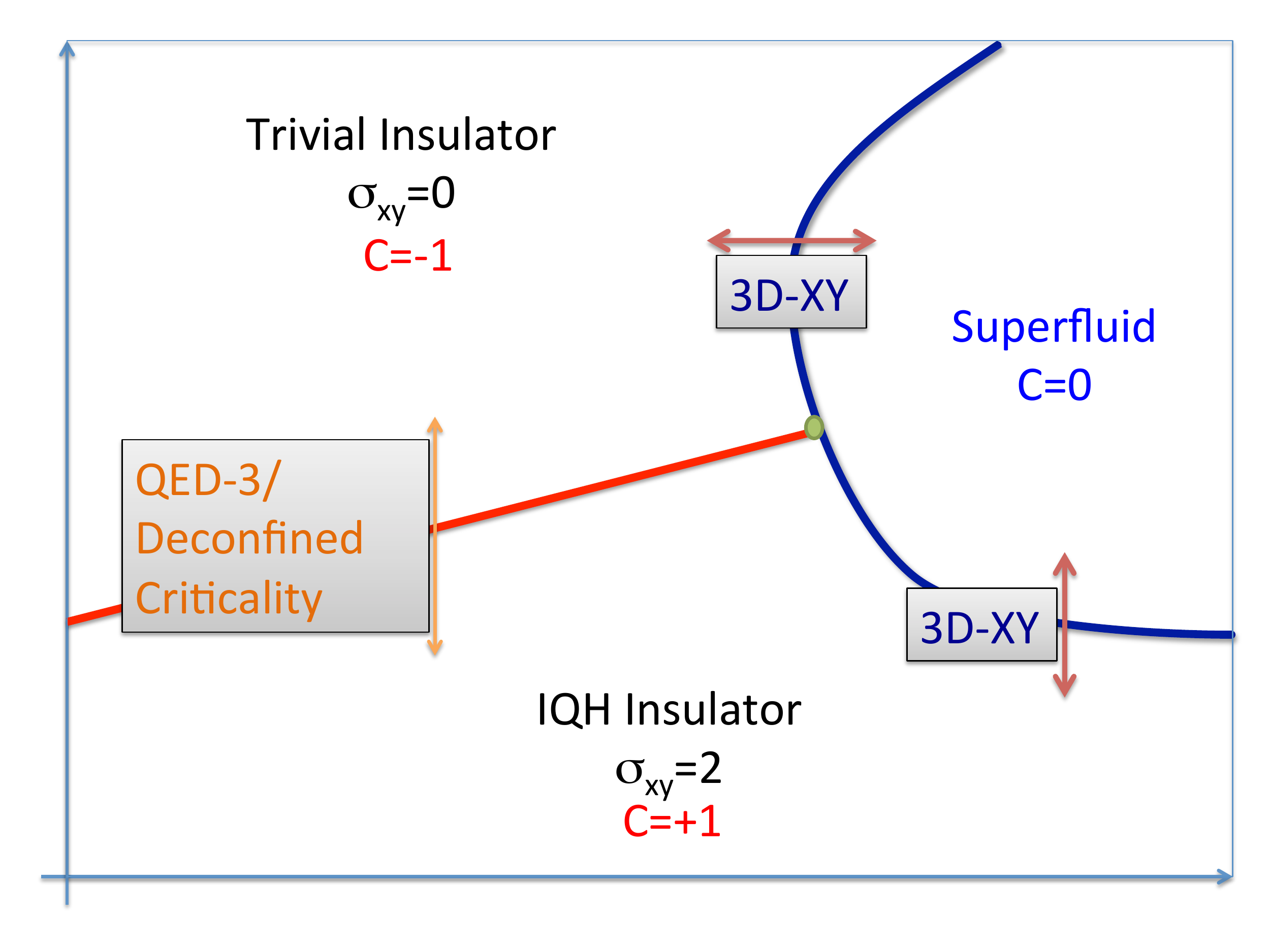}}
\caption{The generic phase diagram in the presence of additional symmetries, inversion and charge conjugation. Direct transitions between any two of the three phases - IQH insulator, trivial insulator and superfluid are all potentially allowed. Here, $C$ denotes Chern number of the parton $f_2$. Changing $C=0 \rightarrow C=1$ ($C=0 \rightarrow C=-1$), corresponds to the  transition between the superfluid and bosonic IQH  insulator (superfluid and trivial insulator) that lies in 3D XY universality class.  
However, when $C= -1\rightarrow C=+1$ describes the critical point between the trivial and IQH insulator,  described by QED-3 with $N_f=2$ species of gapless fermions. This transition is the main subject of this paper.  }
\label{fig:phasediainv}
\end{figure}

As one notices from Table \ref{table1}, a direct phase transition between the SPT phases requires a change in the Chern number for the parton $f_2$ from -1 to 1. Generically, this will require additional lattice symmetries which we specify below. In their absence, the Chern number of $f_2$ changes from -1 to 0 as one approaches from the trivial Mott phase and 1 to 0 as one approaches from the non-trivial Mott phase. The intermediate phase with Chern number of 0 for the $f_2$ parton corresponds to a superfluid and thus we obtain the generic phase diagram shown in Fig.\ref{fig:phasedianosym}.  Below, we first consider the interesting possibility that certain additional symmetries allow for a change in Chern number by two for the $f_2$ parton leading to a non-Landau transition between the SPT phases (Fig. \ref{fig:phasediainv}). As we will see, the additional symmetries required are lattice inversion and charge conjugation symmetry.

\subsection{Direct Phase transition between bosonic IQH Phases} \label{sec:direct}
In order to achieve a direct transition between the trivial insulator and the bosonic IQH state with $\tilde{\sigma}_{xy}=2$, we need the Chern number of the fermion $f_2$ to change from $C=-1$ to $C=+1$. This is achieved if there are a pair of massless Dirac points present at the transition which both acquire a quantum Hall mass term of the same sign. Generically, the band gap in an insulator closes at a single point in the Brillouin zone. However, in the presence of inversion symmetry one can arrange for a pair of Dirac nodes at opposite momenta in the Brillouin zone to be related by symmetry and accomplish this transition (A similar device was used in Ref. \onlinecite{maissam2012} in a  different context). In the following, we denote the low lying fields for the parton $f_2$ near the transition by $f_{2+}$ and $f_{2-}$ where $+/-$ labels node index.

Before proceeding, it is helpful to re-express the low-energy theory for the two phases
in terms of the internal  gauge fields $a_i$ (since the Chern number changing field $f_2$ is charged under $a_3$). This can be achieved by integrating out the gauge fields $\alpha^I$ in Eqns.\ref{nontrivL1} 
and \ref{trivL1}. Performing this exercise results in the following Lagrangian for the two phases \cite{footnotecurrent}:

\bea
\mathcal{L}_{trivial} & = & \frac{1}{4\pi}\epsilon \left[ a_1 \, \partial a_1 + 2 a_1 \, \partial a_2 + 4 a_1 \, \partial A_c \right. \nonumber \\
& &  + 2 a_2 \, \partial a_3 + 2 a_2 \, \partial A_c  + 2 A_c \, \partial A_c \nonumber \\
& & \left. - 2 a_3 \, \partial a_3 \right]
\eea
and 
\bea
\mathcal{L}_{topological} & = & \mathcal{L}_{trivial} +  \frac{2}{4\pi}\epsilon a_3 \, \partial a_3 \nonumber
\eea

Observing that $\mathcal{L}_{trivial}$ and  $\mathcal{L}_{topological}$ differ from each other only in  the term $\frac{2}{4\pi} a_3 \, d a_3$. One
can now write down the critical theory:
\bea
\mathcal{L}_{critical} & = &  \sum_{\alpha = +/-} \overline{f}_{2,\alpha} [\gamma_\mu(-i \partial_\mu  - a_{3\mu} ) ] f_{2,\alpha} \nonumber \\&&+ \frac{1}{4\pi} \epsilon a_3 \, \partial a_3  + \mathcal{L}_{trivial}    
\eea

\begin{table*}
\begin{tabular}{ |c|c | c | c| c|c|c|}
    \thickhline          
\textbf{Symmetry} & $b_\upa(\bf{r})$ & $b_\dna(\bf{r})$ & $f_+(\bf{r})$ & $f_-(\bf{r})$ & $ z_\upa(\bf{r})$ & $z_\dna(\bf{r})$\\ \thickhline
Number Conservation, $U(1)$ & $e^{i\phi} b_\upa(\bf{r})$ & $e^{i\phi} b_\dna(\bf{r})$ & $e^{i\phi} f_+(\bf{r})$ & $e^{i\phi} f_-(\bf{r})$ & $e^{i\phi} z_\upa(\bf{r})$ & $e^{i\phi} z_\dna(\bf{r})$\\ \hline
Translational: $\bf{r} \rightarrow \bf{r'}= \bf{r+a}$, $U(1)$ & $ b_\upa(\bf{r'})$ & $b_\dna(\bf{r'})$ & $e^{i\theta} f_+(\bf{r'})$ & $e^{-i\theta} f_-(\bf{r'})$ & $ e^{i\theta}z_\upa(\bf{r'})$ & $e^{-i\theta} z_\dna(\bf{r'})$\\ \hline
Inversion: $\bf{r} \rightarrow \bf{r'}= -\bf{r}$, $\mathbb{Z}_2$ & $b_\upa(\bf{r'})$ & $b_\dna(\bf{r'})$ & $\gamma_0 f_-(\bf{r'})$ & $\gamma_0f_+(\bf{r'})$ & $ z_\dna(\bf{r'})$ & $z_\upa(\bf{r'})$ \\ \hline
Charge Conjugation, $\mathbb{Z}_2$ & $b^\dagger_\upa(\bf{r})$ & $b^\dagger_\dna(\bf{r})$ & $f^\dagger_-(\bf{r})$ & $f^\dagger_+(\bf{r})$ & $ z^*_\dna(\bf{r})$ & $z^*_\upa(\bf{r})$\\ \hline
\end{tabular}
\caption{Action of physical symmetries on the bosons $b_\upa, b_\dna$, fermionic partons $f_{+}, f_{-}$ and the complex bosons $z_\upa, z_\dna$.}
\label{table2}
\end{table*}

Integrating out $f_2$ with one sign of the mass induces the term  $-\frac{1}{4\pi} \epsilon a_3 \, d a_3$ leading to a trivial phase, while the same term with opposite sign is generated in the topological phase. 

Integrating out the gauge fields $a_1$ and $a_2$ in favor of $a_3$ one obtains:
\bea
\mathcal{L}_{critical}& = &  \sum_{\alpha = +/-} \overline{f}_{\alpha} [\gamma_\mu(-i \partial_\mu  - a_{3\mu} ) ] f_{\alpha}  \\+& & \frac{(\partial_\mu a_{3\nu}-\partial_{\nu}a_{3\mu})^2 }{2g^2}
  - \frac{1}{2\pi}\epsilon A_c \partial a_3  -  \frac{1}{4\pi} \epsilon A_c{\partial A_c}   \nonumber
\label{qed3}
\eea
where we have suppressed the index 2 for the $f$ fermions. This is our main result. Note, there is complete cancellation of the Chern-Simons coefficient for the $a_3$ field and the lowest order term for $a_3$ is a Maxwell term.
Crucially, Eqn.\ref{qed3} leads to the following expression for the charge current:
\be
J_c = \frac{\delta \mathcal{L}}{\delta A_c} = -\frac{1}{2\pi} \epsilon (\partial a_3 + \partial A_c)
\label{eq:current}
\ee

which implies that the physical boson current can be identified with $-\frac{\epsilon_{\mu \nu \lambda} \partial_\nu a_{3\lambda}}{2\pi}$. Since the total boson number is conserved, therefore,  \textit{the flux of the gauge field $a_3$ is conserved at the phase transition}. Hence monopoles are absent, and it corresponds to a {\em non-compact} gauge field. Thus, Eqn. \ref{qed3}, in the absence of external probe field $A_c$, is the Lagrangian for non-compact QED-3 coupled to two flavors of fermions.

{\bf Fermionic Vortices:} Furthermore, we observe that the $f$ fermions are minimally coupled to the gauge field $a_3$ in Eqn.\ref{qed3} . Following the standard boson-vortex duality \cite{zee}, this implies that the fermions are \textit{vortices} of the bosonic degrees of freedom. The vortex current $j_v$ can be identified with $\overline{f} \gamma_\mu f$. This provides an intuitive picture of the transition. Note that the photon of the gauge field is simply the Goldstone mode of the superfluid. Gapping this photon corresponds to realizing an insulating phase where the U(1) symmetry is restored. In D=2+1, there are two ways to provide a gap to the photon, via the Higgs mechanism, or by inducing a Chern Simons term for the gauge field. Confinement is not an option since monopoles are forbidden. The Higgs mechanism is the usual way to realize an insulator and assumes bosonic statistics for vortices that condense at the transition \cite{FisherLee}. However, if the vortices are fermionic, then one can naturally realize a Chern Simons term by inducing a Chern number in the ground state of these vortices. Indeed, this corresponds to putting $f_2$ fermions in a band with non-zero Chern number. Furthermore, it is essential that the last term in Eqn.\ref{qed3}, that corresponds to the Chern-Simons term for the $A_c$ probe field, is present. This term ensures that the Hall conductance is quantized to even integers. We will later discuss the surface of a 3D topological insulator with additional time reversal symmetry where such a term is forbidden, and the half quantized Hall effect that results as a consequence.

{\bf Additional Symmetries:} The average flux density $\Phi_a$ of the gauge field $a_3$ is not necessarily zero and  depends on the background charge density $\rho_c$. Therefore, in general a chemical potential $\mu$ that couples to $\epsilon \partial a_3$ is present in Eqn.\ref{qed3}. However, in the presence of an additional $\mathbb{Z}_2$ `charge conjugation' symmetry, one may set the chemical potential $\mu$ to zero. Specifically, consider enforcing the symmetry $\theta \rightarrow -\theta$  and $n\rightarrow -n$ where $[n,\theta] = i$ are the conjugate number and phases for the bosons. Physically, these symmetries arise in spin systems with rotation symmetry about (say) the $z$ axis.   Then,   the number density is:  $S^z \sim n$, while the phase $S_x + i S_y = e^{i \theta}$. In this setup,  $\theta \rightarrow -\theta$ corresponds to rotation by $\pi$ along the $x$ axis and $n \rightarrow -n$ since $S_z \rightarrow -S_z$. With this symmetry, $a_{3\mu} \rightarrow -a_{3\mu}$, which pins chemical potential $\mu$ to zero. This symmetry also rules out a term proportional to $\overline{f}\gamma_\mu f$ in the action. Physically, $ \overline{f}\gamma_0f  = f^\dagger f$ corresponds to vortex density $j_v = \frac{\vec{\nabla} \times \vec{\nabla}\theta}{2\pi}$ for the bosons, which clearly changes sign under $\theta \rightarrow -\theta$. Similarly, the vortex current $ \overline{f}\gamma_i f $ also changes sign under the same symmetry to preserve the continuity equation for vortices. Thus, this symmetry acts like charge conjugation on the vortices $f_2$. Therefore, the total symmetry of our system is $U(1)_{\textrm{(boson number)}} \ltimes \mathbb{Z}_{2 (\textrm{Charge Conugation }) } \times \textrm{lattice inversion}$. \footnote{The internal symmetry of $U(1)\ltimes \mathbb{Z}_{2}$ is\cite{chen2011}  $\mathbb{Z} \times \mathbb{Z}_2$, which implies additional topological phases.  However, for our purposes the states corresponding to bosonic IQH states  remain distinct as before.}

For Dirac nodes at incommensurate wavevectors, the lattice inversion symmetry leads to a pair of Dirac points at different wave vectors, which effectively leads enlarges the discrete translational symmetry to a continuous $U(1)$ symmetry, if the relative wave vector is incommensurate with the underlying lattice. As written in Eqn.\ref{qed3}, the critical action has $SU(2) \times U(1)$ continuous symmetry. Thus it needs to be supplemented with terms such as $\Delta \mathcal{L} \propto (\overline{f} \sigma_z f)^2$ that break the flavor $SU(2)$ symmetry down to $U(1)$.

\subsection{Universal Properties of the Transition} \label{sec:universal}
The critical theory between the trivial and non-trivial Integer quantum Hall states of bosons has the form of 2+1-d  quantum electrodynamics (Eqn.\ref{qed3}). This immediately implies that the critical theory is a conformal field theory and has a dynamical critical exponent $z=1$. This theory is amenable to a large-$N_f$ expansion where $N_f$ is the number of fermion flavors. 

{\bf Transport Properties:}
At the critical point between the IQH insulators, the charge gap closes and leads to a metal with finite $\sigma_{xx}$ and $\sigma_{xy}$. These can be readily studied in the limit of $T=0$  \cite{SachdevBook}. Let us assume that the fermions have a universal conductance $\sigma_f=\frac1{2\pi}\tilde{\sigma}_f$ in the QED-3 theory at $N_f=2$. Now, since these fermions are vortices, we will show that the universal charge conductance of the bosonic IQH transition  is $\tilde{\sigma}_{xx}= \frac1{\tilde{\sigma}_f}$. Also, $\tilde{\sigma}_{xy}=1$.

To see this we will introduce an external electromagnetic potential that induces an electric field $\epsilon \partial A_c = (0,\,-E_y,\,E_x)$, and calculate the resulting charge current in Eqn. \ref{eq:current}, the relevant components being:
\begin{equation}
J_x =\frac1{2\pi}e_y +\frac{1}{2\pi}E_y;\,\,\, J_y =-\frac1{2\pi}e_x -\frac{1}{2\pi}E_x
\label{Ohmslaw}
\end{equation}
where $\epsilon \partial \langle a_3\rangle =(b,\,-e_y,\,e_x)$ is the internal gauge field strengths induced by the applied field. This is evaluated by noting that the external field couples to the internal gauge field like a current $\mathcal J=-\epsilon \partial A/2\pi$  in the third term of Eqn. \ref{qed3}. Thus, ${\mathcal J}_x= E_y/2\pi$ and ${\mathcal J}_y= -E_x/2\pi$. These currents can be related to the induced internal gauge fields via the resistivity of the fermions in QED-3, $e_a=\rho_f {\mathcal J}_a$. Thus:
\begin{equation}
e_x =-\frac{\rho_f}{2\pi}E_y ;\,\,\,\, e_y =\frac{\rho_f}{2\pi}E_x 
\end{equation}
Substituting this into  Eqn. \ref{Ohmslaw}, and using  the definition $\rho_f=2\pi/\tilde{\sigma}_f$, we finally arrive at the following physical conductivity tensor:
\begin{equation}
{\bf \tilde{\sigma}}=2\pi{\bf \sigma} = \ \begin{bmatrix} {1/\tilde{\sigma}_f} & 1 \\ -1 & {1/\tilde{\sigma}_f} \end{bmatrix}
\end{equation}
thus the universal conductance at the transition is equal to the universal \textit{resistance} of fermions in QED-3.

{\bf Large-N limit:} For illustration, let us explicitly compute these in the limit of large number of flavors $N_f$ for the partons $f_2$. An appropriate large-N generalization of the critical QED-3 theory is:
\bea
\mathcal{L}_{N}& = &  \sum_{\alpha = 1}^{N_f} \overline{f}_{\alpha} [\gamma_\mu(-i \partial_\mu  - a_{\mu} ) ] f_{\alpha}  + \frac{N_f(\partial_\mu a_{\nu}-\partial_{\nu}a_{\mu})^2 }{2g^2}\nonumber
\eea
The conductivity of fermions is readily determined in the large $N_f$ limit where gauge fluctuations can be ignored, yielding $\tilde{\sigma}_f=N_f\sigma_0$ where ${\sigma}_0=\frac{\pi}{8}$\cite{ludwig1994}. The physical conductivity then is $\tilde{\sigma}_{xx}=\frac{1}{N_f\sigma_0}$ which can be estimated at $N_f=2$.  It is amusing to compare this with the clean {\em fermionic} IQH transition where the universal conductivity at the transition is $\tilde{\sigma}^{\rm fIQH}_{xx}=\sigma_0$ and $\tilde{\sigma}^{\rm fIQH}_{xy}=1/2$\cite{ludwig1994}. 

{\bf Critical Exponents:} To begin with, consider the correlations of the boson creation operator $b$ at the transition. Recall that the boson density $b^\dagger b $ is given by the flux operator $\frac{\nabla \times a_3}{2\pi}$ (Eqn.\ref{eq:current}). Therefore, the boson creation operator $b^\dagger$ is identified with the single monopole creation operator for the gauge field $a_3$. Following Borokhov et al \cite{borokhov2002}, the scaling dimension of the monopole operator in QED-3 for $N_f$ flavors of fermions is given by $\Delta_{monopole} =0.265 N_f$ at the leading order in $N_f$. This implies that in a $1/N_f$ expansion, at $N_f = 2$, (assuming the continuous transition persists down to  $N_f=2$) the correlation function of the bosons is given by
\bea
\left< b^\dagger(r) b(r') \right> \sim \frac{1}{|r-r'|^{1+\eta}}
\eea
where $\eta \approx 0.06$.

The energy operator $|b_\upa|^2 + |b_\dna|^2$  in the boson language translates to the mass term  $\overline{f}f$ for the fermionic partons. The scaling dimension of this operator $\Delta_{\overline{\psi}\psi}$ is directly related to the correlation length critical exponent $\nu$ via $\nu^{-1} = 3-\Delta_{\overline{\psi}\psi}$ and is given by \cite{hermele2005}  
$\nu^{-1} = 1- \frac{128}{3\pi^2 N_f}. $ At $N_f = \infty$, one finds $\nu = 1$. However, the leading  $1/N_f$ correction  leads to a negative value for $\nu$ which implies that higher order corrections are important for  $\Delta_{\overline{\psi}\psi}$.

{\bf Possible Connections to Bosonic Deconfined Criticality:}  With these symmetries, the above action can be bosonized \cite{senthil2006} to yield an 
$O(4)$ model with topological $\theta$-term at $\theta = \pi$ with $U(1) \times U(1)$ anisotropy for two complex bosons $z_\upa, z_\dna$ (see Appendix A):

\be 
\mathcal{L} = \frac{1}{g}\left(|(-i\partial_\mu -A_{c\mu}) z_\upa |^2 +  |(-i\partial_\mu-A_{c\mu}) z_\dna|^2\right) + i\pi H
 \label{eq:O4theta}
\ee
The two bosons $z_\upa, z_\dna$ together form a four component real vector $V = [\textrm{Re}(z_\upa)\,\, \textrm{Im}(z_\upa) \,\,\textrm{Re}(z_\dna)\,\, \textrm{Im}(z_\dna)]$. $H$ is the theta term for the vector $V$ which counts the integer associated with the winding number of vector $V$ in space-time. Again, the above action needs to be supplemented with terms such as $\Delta \mathcal{L} \propto (|z_\upa|^2 - |z_\dna|^2)^2$, that produce the $U(1) \times U(1)$ continuous symmetry of our system. Note however that the fields entering this O(4) model description transform differently than those that appear in the network model derivation \cite{senthil2012} of this transition. For example, the $z_{\upa,\dna}$ fields transform under translations, in contrast to the network model fields. 

This model is further dual to the easy-plane $\textrm{NCCP}_1$ model \cite{senthil2006}:
\be
\mathcal{L} = \frac{1}{g'}|(-i\partial_\mu - \alpha_\mu) z'|^2 + \frac{(\nabla \times \alpha)^2}{2e^2} \label{nccpdual}
\ee
where $z' = [z'_1\,\,\, z'_2]$ is a two component complex boson with $|z'_1|^2 = |z'_2|^2 = \frac{1}{2}$ while $\alpha$ is a non-compact $U(1)$ gauge field. We review the duality between NCCP$_1$ and QED-3 with two flavors of fermions in Appendix \ref{sec:qedtonccp}. The $NCCP_1$ theory with $SU(2) \times U(1)$ symmetry arises in the context of deconfined quantum critical point between Neel and Valence Bond Solid (VBS) phases and there is good numerical evidence that the square lattice $J-Q$ model realizes $NCCP_1$ critical theory at the phase transition between Neel and VBS \cite{ashvin2003, senthil2004, sandvik2007, melko2008}. The evidence for a continuous phase transition with easy plane anisotropy is less clear \cite{kuklov, wiese}. In passing, we also note that the duality between QED-3 and $O(4)$ model at $\theta = \pi$ bears some resemblance to the recently proposed dualities between bosonic and fermionic Chern-Simons matter theories \cite{aharony}, though we are unable to find an exact correspondence. It will be interesting to explore any possible connection.

It is important to understand how the physical symmetries of our system are implemented in the various formulations described above. The action of symmetries on the original bosons and the fermionic partons is summarized in Table \ref{table2}.  The gauge fields $a_3$ transform as vectors under inversion, and change sign under charge conjugation. We conclude this section with a word of caution in relating the fermionic QED-3 at $N_f=2$ to these bosonic critical points\cite{senthilcommunication}. Even in the absence of chiral symmetry breaking, it is presently unclear if these bosonic theories can describe critical `fermions'. Instead they may only capture transitions between phases where the symmetry protecting the Dirac points is broken.  Nevertheless this potential web of dualities between different models of deconfined criticality is worth noting.

\subsection{Phase Transition between IQH Phases and Superfluid} \label{sec:sptsf}
As mentioned earlier, if the underlying symmetry of the system is just boson number conservation, then  generically there exists an intervening superfluid phase between the two SPT phases (Fig. \ref{fig:phasedianosym}). In the absence of particle-hole symmetry this transition is ordinary Bose-Einstein condensation transition, with mean-field exponents and dynamic critical exponent $z=2$ \cite{SachdevBook}. As pointed out earlier, even in the presence of charge-conjugation and inversion symmetry, generically there will be regions of phase diagram where the SPT phases undergo transition to a superfluid. Owing to the particle-hole symmetry, this phase transition lies in the 3D XY universality class. It is known that 3D XY transition has an alternative description in terms of Dirac fermions coupled to a Chern-Simons gauge field \cite{chen1993} and as we show now that our parton formulation indeed recovers this alternative description, providing a non-trivial check on our construction.  

Let us first consider the transition from the superfluid to the trivial SPT phase. From Table \ref{table1}, the phase transition occurs as the Chern number of $f_2$ changes from 0 to -1. The critical theory is given by:

\bea
\mathcal{L}_{critical} & = &  \overline{f}_{2} [\gamma_\mu(-i \partial_\mu  - a_{3\mu} ) ] f_{2} + \frac{1}{4\pi} \left(- \frac{3 \epsilon a_3 \, d a_3}{2} \right.\nonumber \\
& & + \epsilon a_1 \, \partial a_1 + 2\epsilon a_1 \, d a_2 + 2\epsilon a_2 \, d a_3) \nonumber \\
\eea

Integrating out $a_2,a_3$ leads to

\bea
\mathcal{L}_{critical} & = &  \overline{f}_{2} [\gamma_\mu(-i \partial_\mu  - a_{3\mu} ) ] f_{2} - \frac{1}{4\pi} \left( \frac{\epsilon a_3 \, d a_3}{2} \right)\nonumber\\ 
\label{eq:crit_trivtosf}
\eea

which is exactly the fermionic dual of 3D XY transition as first described by Chen et al \cite{chen1993}.

A similar analysis for the transition between the superfluid and the non-trivial SPT phase yields the same theory as above but with a reversed sign for the Chern Simons term (last term in Eqn. \ref{eq:crit_trivtosf}) 
which is expected to have the same critical properties since they do not depend on the sign of the Chern-Simons coefficient.

\subsection{Relation to Surface States of 3D Bosonic SPT Phases:} \label{3dspt}
 Recently, a physical model of the unusual surface states of bosonic topological phases\cite{chen2011}  in 3+1 dimensions was discussed \cite{senthil2012}. In particular, consider the symmetry $U(1) \times Z^T_2$ where in addition to the $U(1)$ symmetry that leads to charge conservation, $Z^T_2$  time-reversal symmetry is imposed. In Ref. \onlinecite{senthil2012}, it was argued that with these  symmetries there is a 3D topological phase with the following surface properties. If superfluid order is induced on the surface of this topological phase, vortices in this superfluid are fermionic. The consequences of a finite vortex density were also discussed. However, with an additional $\mathbb{Z}_2$ charge conjugation symmetry, vortices will be effectively at zero density and typically gapped. Now the vortex insulator  is described by Eqn\ref{qed3}, although a mass term that mixes fermions at opposite nodes is allowed. However, the last term in Eqn\ref{qed3}, $\epsilon A_c\partial A_c/4\pi$ which breaks time reversal symmetry, is forbidden. That term was crucial in ensuring an even Hall conductance in the insulating phase, when the $f$ fermion band acquires a Chern number. Now, on the surface of the 3D bosonic topological phase, let us break time reversal and allow for a  Chern number change of the $f$ fermion. Now, in the absence of the $\epsilon A_c\partial A_c/4\pi$ term, an odd integer Hall conductance can be realized, which is only allowed because this bosonic system is the boundary of a three dimensional topological phase. Thus it appears that while fermionic vortices can arise in a 2D superfluid which {\em breaks} time reversal symmetry, they are forbidden in a time reversal symmetric superfluid in D=2+1 dimensions. However, they can arise on the surface of a 3D topological insulator of bosons, even with time reversal symmetry. 
This  is similar to the relation between fermionic integer quantum Hall transitions and the surface states of three dimensional topological insulators. The critical theory in this case consists of free Dirac fermion and for gauge invariance, it needs to be supplemented by a Chern-Simons term $\epsilon A_c\partial A_c/8\pi$  that breaks time-reversal. The boundary of the three dimensional topological insulators is also described by a free fermion, though the Chern-Simons term is absent and the theory is time-reversal invariant.

\section{Conclusions} \label{sec:concl}
In this paper we studied quantum phase transitions involving bosonic topological insulators. To describe these phases, we devised a fermionic parton construction that is distinct from the earlier approaches  including the non-linear sigma model approach \cite{chen2011} or the flux attachment picture \citep{levin2012}. The parton approach allows us to describe the phase transition in terms of Chern number changing transitions for partons. The resulting critical point is described in terms of fermionic vortices that are coupled to internal gauge fields. 

Are bosonic SPT phases as `hard' to realize as states with topological order and fractionalization? A more precise question is to study the evolution of the system from a trivial to a topological phase, and ask if exotic excitations emerge on the way. Indeed, for the bosonic QH states, if symmetry is preserved, we expect  a fermionic version of deconfined criticality of frustrated magnets \cite{senthil2004}, if there is a continuous transition. If however the symmetry is broken, via an intermediate superfluid phase, then all transitions are conventional.

It is interesting to contrast the SPT phase transition presented in this paper with the transitions involving  fermionic integer quantum Hall states. In a clean system, two integer quantum Hall states will generically be separated by an intervening metal, similar to the superfluid phase that separates the two SPT phases in our case. Since any amount of disorder localizes the metal in two dimensions, this leads to the disappearance of the intermediate metal and leads to a direct transition between the integer Hall states \cite{ludwig1994}. In contrast, in our problem, the superfluid phase is stable to weak disorder and therefore our generic phase diagram is expected to remain stable for small disorder. From this point of view, it will be interesting to consider the effect of strong disorder ion bosonic IQH phase transitions.  An alternative approach to fermionic plateau transitions, which does lead to a direct transition without an intervening metal in the clean limit, is to turn on a periodic potential \cite{wen1993, chen1993, sachdev1998, wen2000}. Formally, our approach to bosonic plateau transitions is similar to this latter approach. One can ask if there is an analog of the disordered fermionic quantum critical point that describes the plateau transition, here in the bosonic case. This is an interesting but rather challenging problem, which is left for future work.

The fermionic parton construction for the SPT phases described in this paper can be used to construct mean-field theories, as well as variational wavefunctions for specific bosonic Hamiltonians to search for the presence of SPT phases. The parton construction for SPT phases is rather different from the usual slave-particle techniques since it does not lead to any fractionalized excitations in the gapped phase. It will be worthwhile to generalize it to other SPT phases in both two and three dimensions.

\textit{Acknowledgments}: We thank Leon Balents and M. Metlitski for useful discussions. AV is grateful to T. Senthil for several discussions and a recent collaboration and acknowledges support from NSF DMR-1206728. This research was supported in part by NSF PHY11-25915. On completing this work we learnt that  Y. M. Lu and D. H. Lee have arrived at similar conclusions.

\appendix

\section{Review of the Duality between QED-3 and $O(4)$ Model at $\theta = \pi$}  \label{sec:qedtonccp}
 
Let us start with the QED-3 action (Eqn.\ref{qed3}), and bosonize it by first coupling $\overline{f}_2 \vec{\sigma} f_2$ to a three component vector $\vec{N}$ and then integrating out the fermions ($\vec{\sigma}$
acts on the nodal index +/-). This leads to :

\bea
S & = & \int d^3x\,\, \frac{1}{g} (\partial \vec{N})^2 + i J^T_\mu a_{3\mu} - \frac{1}{2\pi}\epsilon a_3  \partial A_c - i \pi H[\vec{N}] \nonumber \\
\eea  

where $H[\vec{N}]$ is the Hopf term at $\theta = \pi$. The above action can be further simplified by going to the $CP_1$ representation for the vector $\vec{N} = z^{\dagger} \vec{\sigma} z$. This introduces 
a $U(1)$ gauge redundancy which is taken into account by a $U(1)$ gauge field $a$ that couples minimally to the bosons $z = [z_\upa\,\,z_\dna]^T$. In $CP_1$ representation:

\be
i J^T_\mu a_{3\mu} = \frac{1}{2\pi} \epsilon a_3 \, \partial a
\ee

Therefore, the action becomes

\bea
S & = & \int d^3x\,\, \frac{1}{g} |(-i\partial_\mu - a_\mu) z|^2 +  \frac{1}{2\pi} \epsilon a_3 \, \partial (a-A_c)  \nonumber \\
& & + \textrm{Hopf term} \nonumber
\eea

Integrating out the $a_3$ field generates Meissner term for the combination $a-A_c$ and we can thus set $a=A_c$. The Hopf term is equal to the Pontryagin index for the four-component vector 
$V = [\textrm{Real}(z_\upa)\,\, \textrm{Imag}(z_\upa)\,\, \textrm{Real}(z_\dna)\,\, \textrm{Imag}(z_\dna)]^T$ for the mapping from the space-time to $V$ ($\pi_3(S^3) = \mathbb{Z}$). Due to this, the action becomes $O(4)$ model at $\theta = \pi$ \cite{senthil2006,footnotehopf}:
\be 
\mathcal{L} = \frac{1}{g}\left(|(-i\partial_\mu-A_{c\mu}) z_\upa |^2 +  |(-i\partial_\mu-A_{c\mu}) z_\dna|^2\right) + i \pi H
\ee
where $H$ denotes the space-time winding number of the vector $V$. Since the original model has only $U(1) \times U(1)$ symmetry, 
the action needs to be supplemented by terms that break the $O(4)$ symmetry down to $U(1) \times U(1)$.

\section{Lattice Model of Parton Band Structure with Inversion Symmetry, and Resulting Phase Diagram } \label{sec:lattice}
\begin{figure}[tb]
\centerline{
\includegraphics[width=270pt, height=200pt]{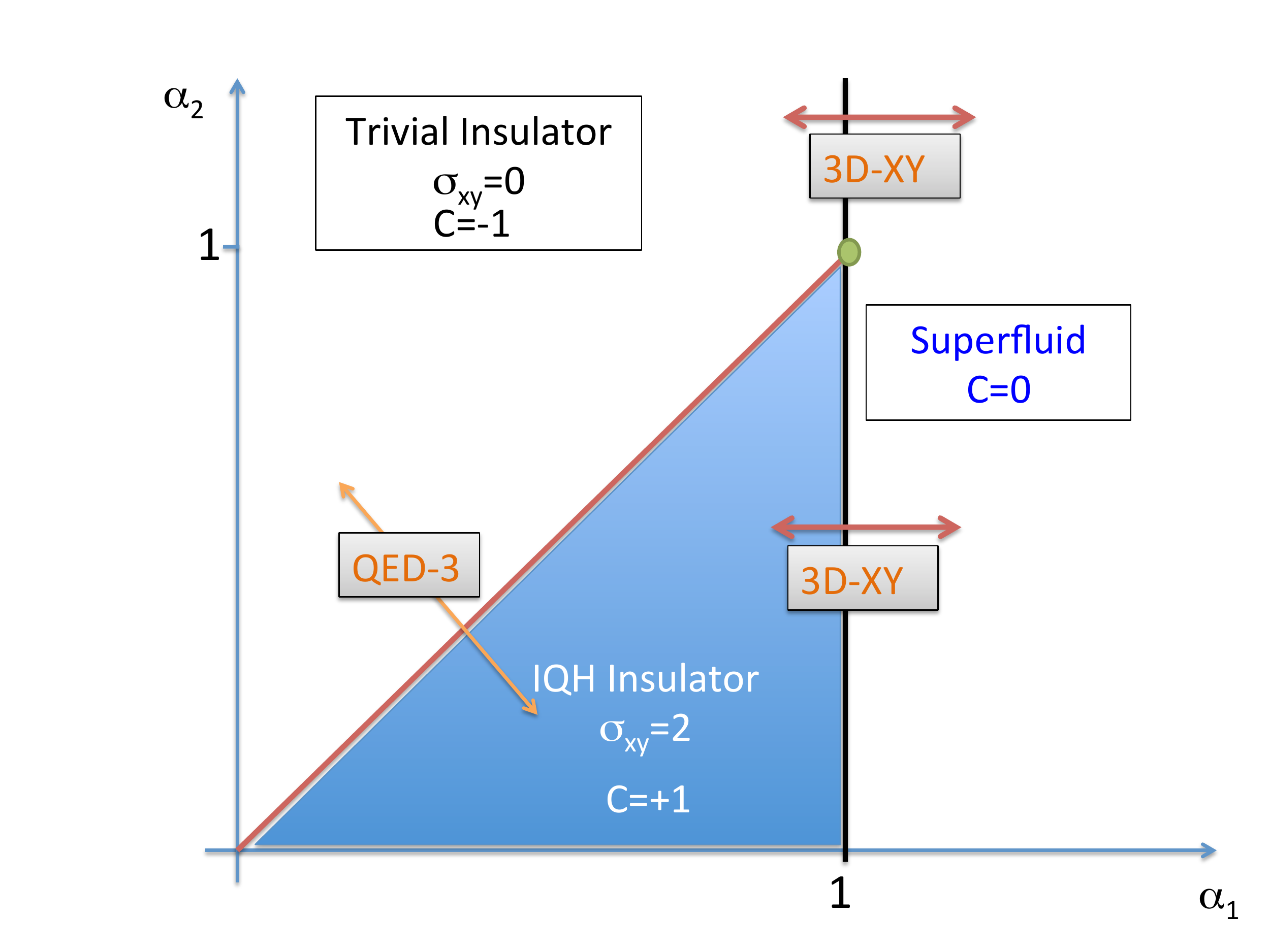}}
\caption{The generic phase diagram for the problem with the following dispersion for the parton $f_2$ as a function of two parameters $a_1,\,a_2$:  $H= \tau_x \sin(k_x) (a_2 - \cos(k_x)) + \tau_y \sin(k_y) + 
\tau_z [(1+ a_1 -\cos(k_x) -\cos(k_y)) ] $. Here, $C$ denotes Chern number of the parton $f_2$. Along the line $a_1 = 1$ and for $a_2 < 1$,  ($a_2>1$) $C=0$ changes  to $C=1$ ($C=-1$) and describes the quantum phase transition between the superfluid and bosonic IQH  insulator (trivial insulator) that lies in 3D XY universality class.  
Finally, and perhaps most interestingly, along the line $a_1 = a_2$, for $a_1 <1$, $C= -1\rightarrow C=+1$ and describes the deconfined critical point between the trivial and IQH insulator, that is equivalently described by QED-3 with $N_f=2$ species of gapless fermions. }
\label{fig:phasediaeg}
\end{figure}

The phase transition between the two SPT phases described above by the $K$-matrices $K = \begin{bmatrix} 0 & 1 \\ 1 & 0 \end{bmatrix}$ and $K = \begin{bmatrix} 0 & 1 \\ 1 & 2 \end{bmatrix}$  occurs when the parton $f_2$ becomes gapless and changes its Chern number from 1 
to -1. One way to conceive such a phase transition is to consider the bosons $b_\upa, b_\dna$ on a lattice with inversion symmetry. Consider, for example, the following two-band Hamiltonian for the parton $f_2$:

\bea 
H & = & \tau_x \sin(k_x) (\alpha_2 - \cos(k_x)) + \tau_y \sin(k_y) + \nonumber \\
& &  \tau_z [(1+ \alpha_1 -\cos(k_x) -\cos(k_y)) ] 
\eea

The Pauli matrices $\tau$ acts on the band index. We note that $H$ is inversion symmetric: $H(\vec{k}) = \tau_z H(-\vec{k}) \tau_z $. The above band structure results in the phase diagram shown in the Fig.\ref{fig:phasediaeg}. Due to the inversion symmetry, along the line $\alpha_1 = \alpha_2$, there are two Dirac nodes at $\pm \vec{k_0}$ where $\vec{k}_0 = (\cos^{-1}(\alpha_1),0)$ which results in the change of the Chern number from -1 to +1, and thereby, phase transition between the trivial Mott insulator and the IQH insulator. To obtain the low-energy theory along the line $\alpha_1 = \alpha_2 \equiv \alpha$, ($\alpha <1$), we write $\vec{k} = \pm \vec{k}_0 + \delta  \vec{k}$ and expand the Hamiltonian $H$ around the two Dirac nodes $\pm \vec{k}_0$. Denoting the parton fields near these two nodes as $\psi_{+}$ and $\psi_{-}$, one finds

\bea
H & = & \sum_{\vec{k}}\psi^{\dagger}_+ \left[\tau_x (1-\alpha^2) \delta k_x + \tau_y \delta k_y + \tau_z \sqrt{1-\alpha^2} \delta k_x \right] \psi_+ \nonumber \\
& & + \psi^{\dagger}_- \left[ \tau_x (1-\alpha^2) \delta k_x + \tau_y \delta k_y - \tau_z \sqrt{1-\alpha^2} \delta k_x\right] \psi_- \nonumber
\eea

To simplify the above expression, we perform a unitary transformation: $\psi_{+} = e^{i \frac{\theta}{2} \tau_y } e^{i \frac{\pi}{4} \tau_x} f_{+}$ and  $\psi_{-} = e^{-i \frac{\theta}{2} \tau_y } e^{i \frac{\pi}{4} \tau_x} f_{-}$, with $\tan(\theta) = \frac{1}{\sqrt{1-\alpha^2}}$, so that 

\bea 
H & = & \sum_{\vec{k}} f^\dagger (\tau_x \alpha \delta k_x  + \tau_z \delta k_y)f
\eea

where $f = [f_{+} \,\,\, f_{-}]^T$ and $\alpha = \sqrt{(1-\alpha^2) + (1-\alpha^2)^2}$. The Euclidean action corresponding the above Hamiltonian is

\be 
S = \sum_{m = +,-}\int d^2x d\tau \overline{f}_m (-i\gamma_0 \partial_{\tau} - i \gamma_x \alpha \partial_{x} - i \gamma_y \partial_{y}) f_{m}
\ee

where $\gamma_0 = \tau_y, \gamma_x = -\tau_z, \gamma_y = \tau_x$ are the three-dimensional Dirac gamma matrices and $\overline{f} = f^{\dagger} (i \gamma_0)$.

Recall that $f$ carries gauge charge of the internal gauge field $a_3$. Coupling to $a_3$ results in QED-3 for two flavors of fermions with anisotropic velocity (due to the factor of $\alpha$ in the action):

\bea
S & = & \sum_{\alpha = +,-}\int d^2x d\tau  \overline{f}_\alpha \gamma_\mu (-i\partial_{\mu} -a_{3,\mu}) f_{\alpha} \nonumber \\&&+ \frac{(\partial_\mu a_{3\nu}-\partial_{\nu}a_{3\mu})^2 }{2g^2}
\eea

where $g$ is the gauge coupling between the $f$ fermions and the gauge field $a_3$. The velocity anisotropy $\alpha$ is irrelevant in the large-$N$ expansion for sufficiently large number of flavors $N_f$ of fermions \cite{vafek, hermele2005} and therefore, we do not consider it. Above action, when supplemented with terms that couple the internal gauge field $a_3$ to the probe field $A_c$ is the putative low-energy theory for the transition between two SPT phases (Eqn.\ref{qed3}).
 
In passing, we note that near the multicritical point $\alpha_1 = \alpha_2 = 1$, where the three phases meet, the dispersion for the fermion $f$ is highly anisotropic and non-relativistic: $H(\vec{k}) \approx \tau_x (\delta k_x)^3 + \tau_y \delta k_y + \tau_z ((\delta k_x)^2 + (\delta k_y)^2)  $ and therefore, if the multicritical point is second-order, it is unlikely that it is described by QED-3, atleast within this model.


\begin{thebibliography}{x}
\bibitem{footnotecurrent} Alternatively, these equations may be obtained by coupling partons to their respective gauge fields and then integrating out the
partons. This way of proceeding does not require introducing gauge fields $\alpha^I$. For example,  \bea
\mathcal{L}_{topological} & = & \frac{1}{4\pi} \left[ (a_1 + A_c) \, \partial(a_1 + A_c) \right. \nonumber \\
 & &  + (a_1 + a_2 + A_c) \, \partial(a_1 + a_2 + A_c) \nonumber \\
& & - a_1 \, \partial a_1 - (a_2 - a_3) \, \partial(a_2 - a_3) \nonumber \\
& & \left. + a_3 \, d a_3 \right]
\eea
\bibitem{footnotehopf} One may ask the following question: the Hopf term may also be written as $i \pi H = \frac{1}{4\pi}b db$. From this, seemingly, when one integrates out the $a_3$ gauge field, it also sets Hopf term to zero,
in contradiction with the claim in Ref.\cite{senthil2006} as well as our result. This erroneous conclusion may be resolved by noticing that the Hopf term can be re-written solely in term of the spinons $z$ as a Pontryagin index associated 
with the vector $V = [\textrm{Real}(z_1)\,\, \textrm{Imag}(z_1)\,\, \textrm{Real}(z_2)\,\, \textrm{Imag}(z_2)]^T$.
\end{thebibliography}
\end{document}